\documentclass[aps,preprint,groupedaddress,amsmath,amssymb] 
{revtex4}

\usepackage[dvips]{graphics}
\usepackage{bm}
\begin{document}

  \title{Elongation/Compaction of Giant DNA Caused by Depletion  
Interaction with a  Flexible Polymer}

\author{M. Kojima}
\affiliation{Department of Physics, Graduate School of Science, Kyoto  
University, Kyoto, 606-8502, Japan}
\author{K. Kubo}
\affiliation{Department of Physics, Graduate School of Science, Kyoto  
University, Kyoto, 606-8502, Japan}
\author{K. Yoshikawa}
\affiliation{Department of Physics, Graduate School of Science, Kyoto  
University, Kyoto, 606-8502, Japan}
\email[To whom correspondence should be addressed. Tel:+81-75-753-3812.  Fax:+81-75-753-3779. Email:]{yoshikaw@scphys.kyoto-u.ac.jp}

\date{\today}

\begin{abstract}
Structural   changes in giant DNA induced by  
the addition of  the flexible
polymer PEG   were examined by    
the method of single-DNA observation.   In dilute DNA conditions, individual DNA 
 assumes a compact state via a
discrete coil-globule transition, whereas in concentrated solution, DNA  
molecules exhibit an extended
conformation via macroscopic phase segregation. The    
 long-axis length of the stretched state in DNA is about  
$10^3$ times larger
than that of the compact  state.   Phase segregation at high DNA concentrations 
occurs at lower PEG    
 concentrations than  the compaction at low DNA  
concentrations.
  These opposite changes in the  
conformation of DNA molecule  are interpreted in terms of
the free energy, including   depletion interaction.
\end{abstract}


\maketitle


\section{Introduction}
\label{intro}

\begin{figure}
\includegraphics{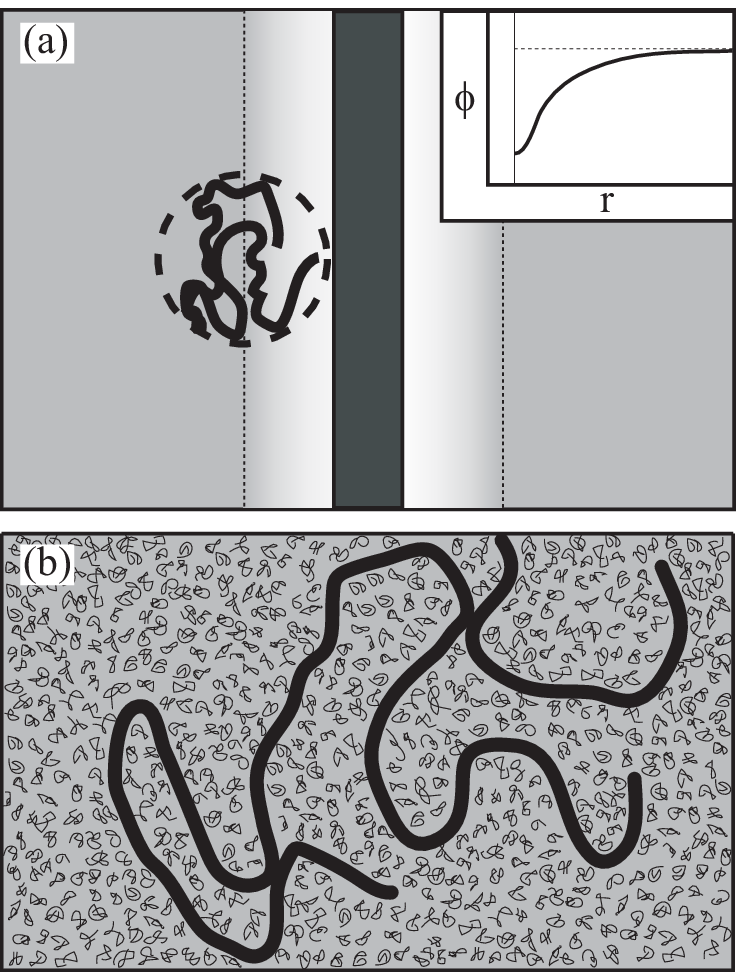}
\caption{}
\label{depletion_diagram}
\end{figure}

It is well  known that semiflexible  polymers  
tend to segregate in the presence of non-adsorbing flexible  
 polymers,
through   so-called depletion  
interaction\cite{Asakura_Oosawa_1954,asakura_oosawa2,Joanny,Lubensky97,Tuinier,Netz_Andelman03}.
 A schematic representation is shown in  
 Fig.\ref{depletion_diagram}.
We consider a flexible polymer  
 that is much smaller than the persistence length\cite{akahon} of  
the semiflexible polymer.
In  this case, the semiflexible polymer 
can be treated as a hard cylinder  at the  
scale of the flexible polymer and
the flexible polymer is depleted from the    
 region of a semiflexible polymer chain(   
 Fig.\ref{depletion_diagram}(a)).
  This region is called the depletion layer.   
When the semiflexible polymer segments come close    
 to each other,
the total depletion layer  decreases and the  
volume occupied by flexible polymers   increases.
  The  presence of the  
flexible polymer causes attractive force   that  
arises from the entropy among the semiflexible polymer  
segments.
 The strength of  
 this attractive force is identical   regardless  
of intra-molecular or inter-molecular interaction
, and only  depends on the steric  
configuration of the interacting segments.
(  Fig.\ref{depletion_diagram}(b)).
 
It has been reported that the formation of  
 a liquid crystalline phase from a concentrated solution  
of short DNA   fragments    
  can be induced by   addition of    
 the flexible polymer PEG, where DNA molecules are extended and  
aligned parallel, accompanied by macroscopic phase  
segregation\cite{Jordan,Livolant,Strey97,osmotic,Harreis02}.
  In contrast to the extension effect by depletion,  
it is also known \cite{Lerman_71,denken,Minagawa,hydrodynamic} that a  
compact state of DNA is generated by the addition of PEG.  Thus,  
 previous reports have indicated   
  opposite effects  of a flexible polymer (either extension or compaction  
of DNA molecules), depending on the experimental conditions on the individual articles.
The purpose of the present study   was to obtain a  
comprehensive view  of macroscopic phase segregation  
and the compaction of a semiflexible polymer by the addition of  
 a flexible polymer.

\section{Experimental}
\label{experiment}
Bacteriophage T4 DNA (166 kbp, contour length 57 $\mu$m) was purchased  
from Nippon Gene (Tokyo, Japan).
The fluorescent dye Quinolinium, 1,19-[1,3-propanediylbis  
[(dimethyliminio)-
3,1-propanediyl]] bis [4-[(3-methyl-2(3H)-benzoxazolylidene)-
methyl]]-, tetraiodide (YOYO-1) was obtained from Molecular  
Probes (Eugene, Oregon).
Polyethylene glycol (PEG) 6000 ( average molecular  
weight $\cong$ 8200) was obtained from Kishida Chemical Co., Ltd.  
(Osaka, Japan).  Analytical-grade NaCl was  
 obtained from Nacalai Tesque (Kyoto, Japan).

All  procedures were performed at room temperature in Tris  
100 mM EDTA 10 mM buffer(TE buffer pH 8.0).
 First, PEG and NaCl were dissolved in a  
microtube.
The  concentration of PEG was adjusted as  
desired.
The final  concentration of NaCl was  
 100 mM in all experiments.
T4 DNA solution was mixed to give  a final  
concentration of 20 $\mu g/ml$ (A)
or 0.1 $\mu g/ml$ (B)  in the microtube.
 Less than 1 \% of T4 DNA in (A)    
 was stained with YOYO-1.  This partially stained  
solution was
mixed with a solution free from dye or solutions    
 that had   already been equilibrated with dye.
 In contrast, the T4 DNA molecules in (B) were  
uniformly stained.
The ratio between labeled DNA and fluorescent dye was [labeled  
DNA(bp)]/[YOYO-1]=5 in molar  units.
It has been confirmed that the contour length and persistence length  
 remain essentially  constant, i.e., the  
effect of dye staining is negligible\cite{yoshinaga}.
The (A) samples  were  gently shaken   and allowed  
to stand   for a few hours. After    
 this rest period, the samples were gently  
vortexed.
The (B) samples were   allowed to stand for 24  
hours after   mixing.

 Fluorescent microscopic images of DNA and  
polarization microscopic images of   
 the same regions
 were observed using a BX 60 (Olympus, Japan)  
fluorescence and polarization   microscope
with a color CCD camera
(Watec, Japan).
The fluorescence images of the DNA chain were observed using an  
Axiovert 200 fluorescence  microscope
(Carl Zeiss, Germany) with a   
 high-resolution EBCCD camera (Hamamatsu, Japan).

\section{Results and Discussion}

\label{r_d}

\begin{figure*}
\includegraphics{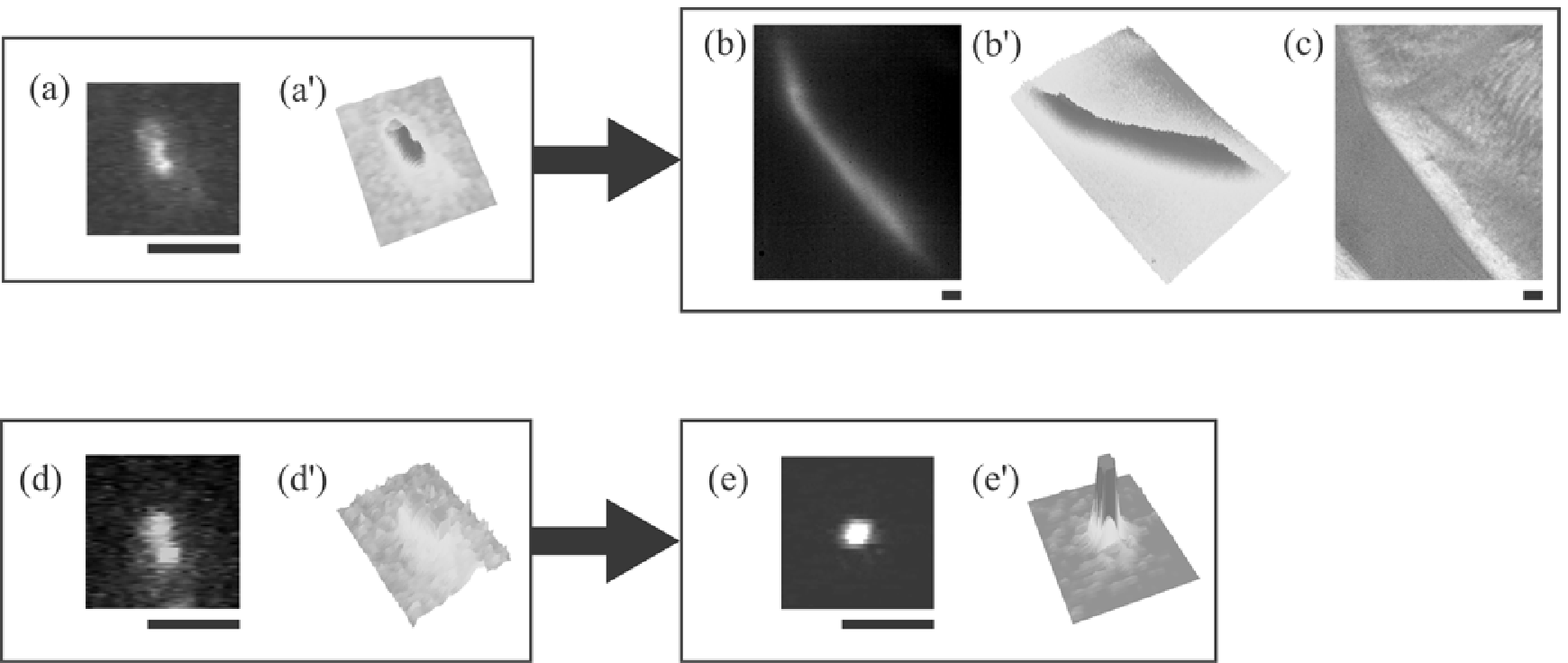}
\caption{}
\label{single_molecule}
\end{figure*}

 Figure \ref{single_molecule} shows typical  
fluorescence images of individual DNA molecules.
At a low PEG concentration, DNA molecules    
 assume a random coil state regardless of the DNA  
concentration (  Fig.\ref{single_molecule}(a), (d)).
With  an increase  in the PEG  
concentration,
DNA  molecules at a high  
DNA concentration  show an elongated conformation  
( Fig.\ref{single_molecule}(b)),
accompanied by  macroscopic phase segregation between  
PEG-rich and DNA-rich  phases.
The  long-axis length of an elongated DNA  
chain approximately corresponds to its contour length;
In  Fig.\ref{single_molecule}(b), the full length is  
about 50 $\mu m$.
 Figure \ref{single_molecule}(c)    
 shows a polarization  microscopy image  
 of the same region as   in  
Fig.\ref{single_molecule}(b).
  The DNA molecule in (B)    
 assumes a compact state through the coil-globule transition  
( Fig.\ref{single_molecule} (e)).
Note that the long axis of an elongated DNA chain is about $10^3$  
times larger than that  in the compact state.

\begin{figure}
\includegraphics{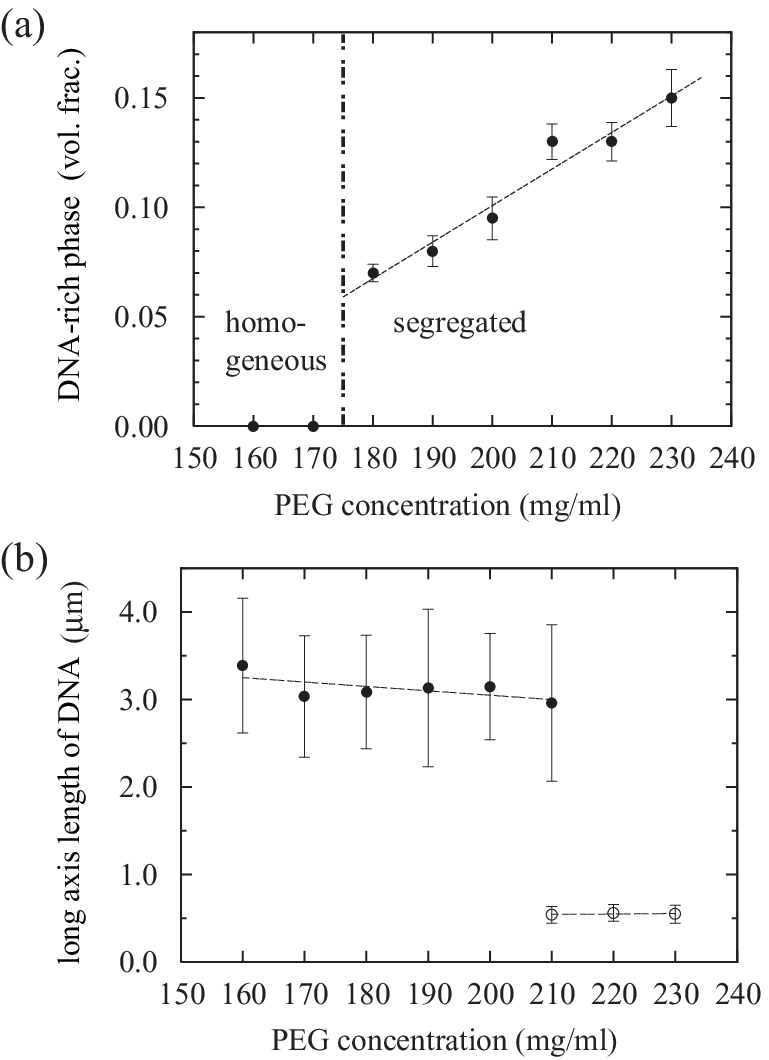}
\caption{}
\label{transitions}
\end{figure}

 Figure \ref{transitions} (a)    
 shows the  dependence of the
volume  fraction of DNA $\phi_{DNA}$ inside the  
generated DNA-rich phase on the PEG concentration.
The lowest PEG concentration  that induced  
phase segregation was 180 mg/ml(  
 Fig.\ref{transitions}(a)).
 The volume fraction of DNA  
$\phi_{DNA}$ inside the generated DNA-rich phase
 appears to increase with  an  
increase  in the PEG concentration.  The line in  
  Fig.\ref{transitions} (a) is fitted by   the least-squares method
with a PEG concentration within the range   of  
180 to 230 mg/ml   and the gradient   is ca. 0.002.
  $\phi_{DNA}$   can be calculated  
by   considering that all of the  
DNA molecules are transferred   to the DNA-rich  
phase.

Figure \ref{transitions} (b) shows the distribution of  
  the long-axis length of DNA   
 molecules through the coil-globule transition
 versus the PEG concentration.
Closed and open circles  show the  
average  long-axis length of DNA  
molecules in the coil and globule states,   respectively.
 DNA molecules in both compact and coiled states   were observed   at a PEG  
concentration of around 210 mg/ml and
the number of DNA molecules in the compact state was  
approximately twice   that  in the coil state.

Note that the critical PEG concentration   for  
phase segregation was lower than that  for the  
coil-globule transition.
This result means that the elongated state of DNA molecules is more  
stable than the compact state if the DNA concentration
is   high enough.

%

Let us discuss  both transitions in terms of free  
energy.
The total free energy  variance per 
DNA chain $\Delta F_{tot}$ can be written as
\begin{equation}
	\Delta F_{tot}\cong\Delta F_{dp} +   \Delta F_{el}
\label{balance}
\end{equation}
where
$\Delta F_{dp}$ describes the gain in the
decrease  in the depletion layer for PEG per   DNA chain.
$\Delta F_{el}$ describes the elastic part of the free energy for a  
single DNA chain.
For simplicity, the electrostatic and van der Waals interactions among  
DNA, PEG and ions are neglected in eq.(\ref{balance}).
  These    
miscellaneous effects are considered to be incorporated  
  in the effective elastic energy.

%
%
${\Delta F_{dp}}$ can be written approximately    
 as\cite{akahon,VVV}
\begin{equation}
	\label{fdp}
	\Delta F_{dp}/kT = \frac{-\Delta V}{d^3}  
(\frac{\phi_{PEG}}{N_{PEG}}\ln \frac{\phi_{PEG}}{N_{PEG}}  
-\frac{\phi_{PEG}}{N_{PEG}}) \hspace{10mm}(<0)
\end{equation}
where T is   temperature, k is the Boltzmann constant,
$\Delta V(<0)$ is the variation in the depletion layer for PEG around  
 a DNA molecule,
$d$ is the elementary spacing in the Flory-Huggins model which was  
defined to
be equal to the width of the DNA Kuhn segment,
$\phi_{PEG}$ is the volume fraction of PEG, and $N_{PEG}$ is the  
  average degree of polymerization of PEG  
  molecules.
In eq.(\ref{fdp}), the translational entropy of water is neglected,
 since the effective volume occupied by water may  
not change   greatly.  We also    
 treat
  $\phi_{PEG}$  as a constant through  
the transitions
because $\Delta V$ is much smaller than the total volume of the system.

$\Delta V$ may be approximately given as\cite{asakura_oosawa2}
\begin{equation}
\label{deltav}
\Delta V \cong \pi L (\frac{d}{2})^2/\phi_{DNA}- \pi  
L(\frac{d}{2}+Rg_{PEG})^2
\end{equation}
where L is the contour length of a DNA chain and $Rg_{PEG} =  
\sqrt{N_{PEG}^{2\nu}b^2 /6}$ is the radius of gyration of a PEG  
molecule.
($\nu$: Flory exponent, $b$: PEG monomer size).
The first term   on the right-hand side is the  
volume occupied by a DNA molecule after the transition.
The second term represents the depletion layer for the center of  
mass of PEG around a DNA molecule before  
 the transition.
For  analytical treatment, the depletion  
layer before the transition is represented as a step  
  function,
although the actual depletion layer will not  be so  
simple because of the continuous PEG concentration profile.

%
%

Following Ref.
\cite{akahon},
the elastic free energy $F_{el}$ asymptotically reads
\begin{eqnarray}
\label{fel}
	\Delta F_{el}/kT&\cong&\frac{3}{2}(\alpha ^2 + \alpha ^{-2})\\
	\alpha &=& \frac{R_{DNA}}{N_{DNA}^{1/2}l}\nonumber
\end{eqnarray}
where $\alpha$ is the expansion factor of the DNA coil with respect to  
its ideal size,
$R_{DNA}$ is the  long-axis length of   
 a DNA molecule, $N_{DNA}$ is the number of Kuhn   
 segments
and $l$ is the Kuhn length of a DNA chain.

In the case of compaction,
$\Delta F_{el}$ can be written   as $\Delta  
F_{tr}$
which is   a toroidal shape of compact  
DNA\cite{Ubbink96}.
\begin{equation}
\label{toroid}
\Delta F_{tr}/kT  \cong  \frac{l L}{(R_{DNA}/2)^2}
\end{equation}

In the present paper, the constants are set as $d$ = 2 $nm$,  
$b$=0.38 $nm$\cite{peg_monomer}, $\nu$=0.6, $N_{PEG}$=186, and  
$L$=57 $\mu m$.
 From eq.(\ref{fdp}), (\ref{fel}) and  (\ref{toroid}),
$\Delta F_{dp}$, $\Delta F_{el}$ and $\Delta F_{tr}$ are calculated as  
in Table \ref{table}.
In the case of   macroscopic phase segregation, $\Delta V$ was  
-1.2$\cdot 10^6 (nm^3)$, the critical PEG concentration was 180 mg/ml  
($\phi_P=0.14$)
and the volume   fraction was $\phi_{DNA} \cong  
0.07$ from our experiment.   Therefore,
$F_{dp}$ was calculated  to be about -900 kT.    
 Since   $R_{DNA}$ was
  about $5\cdot10^4  nm$,
$\Delta F_{el}$ is evaluated to be about +700 kT.
  In  the coil-globule transition,
it   has been determined by measurement of the  
hydrodynamic radius  \cite{hydrodynamic} and a MD  
simulation\cite{noguchi} that the compact state of a DNA molecule  
is completely packed.
Thus, we consider $\phi_{DNA}\cong 1$  to be a good  
approximation.
Thus, $\Delta V$ is calculated as -3.6$\cdot 10^6(nm^3)$ and the  
critical PEG concentration  is 210 mg/ml($\phi_P=0.16$).
$\Delta F_{dp}$ is estimated   to be $-3\cdot 10^3$kT.
 Based on electron-microscopic  
observation\cite{denken}, the   long-axis length  
of toroidal DNA is roughly 70 $nm$.
Consequently, $\Delta F_{el}$ and $\Delta F_{tr}$   
 are deduced  to be +2000 kT and +2300 kT,  
respectively.
  Despite this simple estimation, it  
becomes clear that the gain in free energy involving PEG and the  
loss involving the DNA conformation   are  
 on the same order in both transitions.

In summary, we studied both the segregation  
and   compaction of giant DNA induced by  
 depletion interaction.    
 There is a marked difference in the critical concentrations of  
flexible polymer needed to cause   these  
transitions
and  the essence of   this difference  
 can be interpreted in terms of the change in free  
energy with   depletion interaction.

\begin{acknowledgments}

This work was supported by a Grant-in-Aid for Scientific Research in  
Priority Areas ``System Cell Engineering by Multi-scale Manipulation''
(No. 17076007)
from the Ministry of Education, Culture, Sports, Science  and Technology of Japan.
\end{acknowledgments}


\clearpage

\bibliography{kojima_depletion}

\begin{thebibliography}{21}
\expandafter\ifx\csname natexlab\endcsname\relax\def\natexlab#1{#1}\fi
\expandafter\ifx\csname bibnamefont\endcsname\relax
  \def\bibnamefont#1{#1}\fi
\expandafter\ifx\csname bibfnamefont\endcsname\relax
  \def\bibfnamefont#1{#1}\fi
\expandafter\ifx\csname citenamefont\endcsname\relax
  \def\citenamefont#1{#1}\fi
\expandafter\ifx\csname url\endcsname\relax
  \def\url#1{\texttt{#1}}\fi
\expandafter\ifx\csname urlprefix\endcsname\relax\def\urlprefix{URL }\fi
\providecommand{\bibinfo}[2]{#2}
\providecommand{\eprint}[2][]{\url{#2}}

\bibitem[{\citenamefont{Asakura and F.Oosawa}(1954)}]{Asakura_Oosawa_1954}
\bibinfo{author}{\bibfnamefont{S.}~\bibnamefont{Asakura}} \bibnamefont{and}
  \bibinfo{author}{\bibnamefont{F.Oosawa}}, \bibinfo{journal}{J. Chem. Phys.}
  \textbf{\bibinfo{volume}{33}}, \bibinfo{pages}{183} (\bibinfo{year}{1954}).

\bibitem[{\citenamefont{Asakura and Oosawa}(1958)}]{asakura_oosawa2}
\bibinfo{author}{\bibfnamefont{S.}~\bibnamefont{Asakura}} \bibnamefont{and}
  \bibinfo{author}{\bibfnamefont{F.}~\bibnamefont{Oosawa}},
  \bibinfo{journal}{J. Polym. Sci.} \textbf{\bibinfo{volume}{33}},
  \bibinfo{pages}{183} (\bibinfo{year}{1958}).

\bibitem[{\citenamefont{Joanny et~al.}(1979)\citenamefont{Joanny, Leibler, and
  Gennes}}]{Joanny}
\bibinfo{author}{\bibfnamefont{J.~F.} \bibnamefont{Joanny}},
  \bibinfo{author}{\bibfnamefont{L.}~\bibnamefont{Leibler}}, \bibnamefont{and}
  \bibinfo{author}{\bibfnamefont{P.~G.~D.} \bibnamefont{Gennes}},
  \bibinfo{journal}{J. Polym. Sci. Polym. Phys. Ed.}
  \textbf{\bibinfo{volume}{17}}, \bibinfo{pages}{1073} (\bibinfo{year}{1979}).

\bibitem[{\citenamefont{Lubensky}(1997)}]{Lubensky97}
\bibinfo{author}{\bibfnamefont{T.~C.} \bibnamefont{Lubensky}},
  \bibinfo{journal}{Solid State Communications} \textbf{\bibinfo{volume}{102}},
  \bibinfo{pages}{187} (\bibinfo{year}{1997}).

\bibitem[{\citenamefont{Netz and Andelman}(2003)}]{Netz_Andelman03}
\bibinfo{author}{\bibfnamefont{R.~R.} \bibnamefont{Netz}} \bibnamefont{and}
  \bibinfo{author}{\bibfnamefont{D.}~\bibnamefont{Andelman}},
  \bibinfo{journal}{Phys. Rep.} \textbf{\bibinfo{volume}{380}},
  \bibinfo{pages}{1} (\bibinfo{year}{2003}).

\bibitem[{\citenamefont{Tuinier et~al.}(2003)\citenamefont{Tuinier, Rieger, and
  {de Kruif}}}]{Tuinier}
\bibinfo{author}{\bibfnamefont{R.}~\bibnamefont{Tuinier}},
  \bibinfo{author}{\bibfnamefont{J.}~\bibnamefont{Rieger}}, \bibnamefont{and}
  \bibinfo{author}{\bibfnamefont{C.~G.} \bibnamefont{{de Kruif}}},
  \bibinfo{journal}{Advances in Colloid and Interface Science}
  \textbf{\bibinfo{volume}{103}}, \bibinfo{pages}{1} (\bibinfo{year}{2003}).

\bibitem[{\citenamefont{Grosberg and Khokhlov}(1994)}]{akahon}
\bibinfo{author}{\bibfnamefont{A.~Y.} \bibnamefont{Grosberg}} \bibnamefont{and}
  \bibinfo{author}{\bibfnamefont{A.~R.} \bibnamefont{Khokhlov}},
  \emph{\bibinfo{title}{Statistical Physics of Macromolecules}}
  (\bibinfo{publisher}{AIP PRESS}, \bibinfo{year}{1994}).

\bibitem[{\citenamefont{Jordan et~al.}(1971)\citenamefont{Jordan, Lerman, and
  Venable}}]{Jordan}
\bibinfo{author}{\bibfnamefont{C.~F.} \bibnamefont{Jordan}},
  \bibinfo{author}{\bibfnamefont{L.~S.} \bibnamefont{Lerman}},
  \bibnamefont{and} \bibinfo{author}{\bibfnamefont{J.}~\bibnamefont{Venable}},
  \bibinfo{journal}{Nature New Biology} \textbf{\bibinfo{volume}{236}},
  \bibinfo{pages}{67} (\bibinfo{year}{1971}).

\bibitem[{\citenamefont{Livolant and Leforestier}(1996)}]{Livolant}
\bibinfo{author}{\bibfnamefont{F.}~\bibnamefont{Livolant}} \bibnamefont{and}
  \bibinfo{author}{\bibfnamefont{A.}~\bibnamefont{Leforestier}},
  \bibinfo{journal}{Prog. Polym. Sci.} \textbf{\bibinfo{volume}{21}},
  \bibinfo{pages}{1115} (\bibinfo{year}{1996}).

\bibitem[{\citenamefont{Strey et~al.}(1997)\citenamefont{Strey, Parsegian, and
  Podgornik}}]{Strey97}
\bibinfo{author}{\bibfnamefont{H.~H.} \bibnamefont{Strey}},
  \bibinfo{author}{\bibfnamefont{V.~A.} \bibnamefont{Parsegian}},
  \bibnamefont{and}
  \bibinfo{author}{\bibfnamefont{R.}~\bibnamefont{Podgornik}},
  \bibinfo{journal}{Phys. Rev. Lett} \textbf{\bibinfo{volume}{78}},
  \bibinfo{pages}{895} (\bibinfo{year}{1997}).

\bibitem[{\citenamefont{Leonard et~al.}(2001)\citenamefont{Leonard, Hong, and
  N.Easwar}}]{osmotic}
\bibinfo{author}{\bibfnamefont{M.}~\bibnamefont{Leonard}},
  \bibinfo{author}{\bibfnamefont{H.}~\bibnamefont{Hong}}, \bibnamefont{and}
  \bibinfo{author}{\bibfnamefont{H.~H.~S.} \bibnamefont{N.Easwar}},
  \bibinfo{journal}{Polymer} \textbf{\bibinfo{volume}{42}},
  \bibinfo{pages}{5823} (\bibinfo{year}{2001}).

\bibitem[{\citenamefont{et~al.}(2002)}]{Harreis02}
\bibinfo{author}{\bibfnamefont{H.~M.~H.} \bibnamefont{et~al.}},
  \bibinfo{journal}{Phys. Rev. Lett.} \textbf{\bibinfo{volume}{89}},
  \bibinfo{pages}{018303} (\bibinfo{year}{2002}).

\bibitem[{\citenamefont{Lerman}(1971)}]{Lerman_71}
\bibinfo{author}{\bibfnamefont{L.~S.} \bibnamefont{Lerman}},
  \bibinfo{journal}{Proc. Nat. Acad. Sci. USA} \textbf{\bibinfo{volume}{68}},
  \bibinfo{pages}{1886} (\bibinfo{year}{1971}).

\bibitem[{\citenamefont{Laemnli}(1978)}]{denken}
\bibinfo{author}{\bibfnamefont{U.~K.} \bibnamefont{Laemnli}},
  \bibinfo{journal}{Proc. Nat. Acad. Sci. USA} \textbf{\bibinfo{volume}{72}},
  \bibinfo{pages}{4288} (\bibinfo{year}{1978}).

\bibitem[{\citenamefont{Minagawa et~al.}(1994)\citenamefont{Minagawa,
  Matsuzawa, Yoshikawa, Khokhlov, and Doi}}]{Minagawa}
\bibinfo{author}{\bibfnamefont{K.}~\bibnamefont{Minagawa}},
  \bibinfo{author}{\bibfnamefont{Y.}~\bibnamefont{Matsuzawa}},
  \bibinfo{author}{\bibfnamefont{K.}~\bibnamefont{Yoshikawa}},
  \bibinfo{author}{\bibfnamefont{A.~R.} \bibnamefont{Khokhlov}},
  \bibnamefont{and} \bibinfo{author}{\bibfnamefont{M.}~\bibnamefont{Doi}},
  \bibinfo{journal}{Biopolymers} \textbf{\bibinfo{volume}{34}},
  \bibinfo{pages}{555} (\bibinfo{year}{1994}).

\bibitem[{\citenamefont{Yoshikawa and Matsuzawa}(1995)}]{hydrodynamic}
\bibinfo{author}{\bibfnamefont{K.}~\bibnamefont{Yoshikawa}} \bibnamefont{and}
  \bibinfo{author}{\bibfnamefont{Y.}~\bibnamefont{Matsuzawa}},
  \bibinfo{journal}{Physica D} \textbf{\bibinfo{volume}{84}},
  \bibinfo{pages}{220} (\bibinfo{year}{1995}).

\bibitem[{\citenamefont{Yoshinaga et~al.}(2001)\citenamefont{Yoshinaga,
  Akitaya, and Yoshikawa}}]{yoshinaga}
\bibinfo{author}{\bibfnamefont{N.}~\bibnamefont{Yoshinaga}},
  \bibinfo{author}{\bibfnamefont{T.}~\bibnamefont{Akitaya}}, \bibnamefont{and}
  \bibinfo{author}{\bibfnamefont{K.}~\bibnamefont{Yoshikawa}},
  \bibinfo{journal}{B.B.R.C.} \textbf{\bibinfo{volume}{286}},
  \bibinfo{pages}{264} (\bibinfo{year}{2001}).

\bibitem[{\citenamefont{Vasilevskaya et~al.}(1995)\citenamefont{Vasilevskaya,
  Kohokhlov, Matsuzawa, and Yoshikawa}}]{VVV}
\bibinfo{author}{\bibfnamefont{V.~V.} \bibnamefont{Vasilevskaya}},
  \bibinfo{author}{\bibfnamefont{A.~R.} \bibnamefont{Kohokhlov}},
  \bibinfo{author}{\bibfnamefont{Y.}~\bibnamefont{Matsuzawa}},
  \bibnamefont{and}
  \bibinfo{author}{\bibfnamefont{K.}~\bibnamefont{Yoshikawa}},
  \bibinfo{journal}{JCP} \textbf{\bibinfo{volume}{102}}, \bibinfo{pages}{6595}
  (\bibinfo{year}{1995}).

\bibitem[{\citenamefont{Ubbink and Odjik}(1996)}]{Ubbink96}
\bibinfo{author}{\bibfnamefont{J.}~\bibnamefont{Ubbink}} \bibnamefont{and}
  \bibinfo{author}{\bibfnamefont{T.}~\bibnamefont{Odjik}},
  \bibinfo{journal}{Eruophys. Lett.} \textbf{\bibinfo{volume}{33}},
  \bibinfo{pages}{353} (\bibinfo{year}{1996}).

\bibitem[{\citenamefont{Marsh}(2004)}]{peg_monomer}
\bibinfo{author}{\bibfnamefont{D.}~\bibnamefont{Marsh}},
  \bibinfo{journal}{Biophysical Journal} \textbf{\bibinfo{volume}{86}},
  \bibinfo{pages}{2630} (\bibinfo{year}{2004}).

\bibitem[{\citenamefont{Noguchi et~al.}(1996)\citenamefont{Noguchi, Saito,
  Kidoaki, and Yoshikawa}}]{noguchi}
\bibinfo{author}{\bibfnamefont{H.}~\bibnamefont{Noguchi}},
  \bibinfo{author}{\bibfnamefont{S.}~\bibnamefont{Saito}},
  \bibinfo{author}{\bibfnamefont{S.}~\bibnamefont{Kidoaki}}, \bibnamefont{and}
  \bibinfo{author}{\bibfnamefont{K.}~\bibnamefont{Yoshikawa}},
  \bibinfo{journal}{Chem. Phys. Lett.} \textbf{\bibinfo{volume}{261}},
  \bibinfo{pages}{527} (\bibinfo{year}{1996}).

\end{thebibliography}

\clearpage

\begin{table}
\hspace{10.0cm}
\caption{
\label{table}
Change in free energy accompanied by the transition of coiled DNA into elongated/compact states.
}%
\begin{ruledtabular}
\begin{tabular}{ccccccc}
	& $\Delta V/nm^3$ \footnotemark \footnotetext{volume change in  
 the depletion layer for PEG around a DNA molecule.}\ \
	& $\phi_{PEG}$ \footnotemark \footnotetext{volume fraction of PEG}\ \
	& $\Delta F_{dp}/kT$ \footnotemark \footnotetext{free energy change owe to the depletion interaction per single DNA chain}\ \
	& $R_{DNA}/nm$ \footnotemark \footnotetext{long-axis length of DNA after the transition.}\ \
	& $\Delta F_{el}/kT$ \footnotemark \footnotetext{free energy change of elastic term single DNA chain, estimated from the volume change.}\ \
	& $\Delta F_{tr}/kT$ \footnotemark \footnotetext{free energy change of elastic term by the consideration of toroidal compact state in DNA(Ref.\cite{Ubbink96}.)  
  }\ \
\\
\hline
	elongation  &-1.2$\cdot10^6$ &0.14&  -900 & 5$\cdot$10$ ^4$ & +700 & -  
\\
	compaction &-3.6$\cdot 10^6$ & 0.16 & -3000 &70  \cite{denken} & +2000  
  & +2300\\
\end{tabular}
\end{ruledtabular}
\end{table}

\clearpage

FIG. 1: Schematic representation on the mixed solution of semiflexible and flexible polymers.
(a) Microscopic view on a flexible polymer chain around semiflexible polymer.
Near the surface of the semiflexible polymer, the volume fraction of flexible polymer is lower than that in the bulk.
Inset: Schematic density profile of the flexible polymer,
where $\phi$ is the volume fraction of the flexible polymer and r is the distance from the surface of the semiflexible polymer.
(b) Corse-grained view on the conformation of a semiflexible polymer in the solution of flexible polymer chains. 

FIG.2: Conformational change of single DNA by the addition of PEG.
All scale   bars represent 3 $\mu m$.
(a) Selectively stained DNA molecule with a coiled conformation  
in concentrated DNA solution (DNA 20 $\mu $g/ml, PEG 170 mg/ml, NaCl  
100 mM).
(b) Selectively stained DNA molecule with a stretched  
conformation in a liquid crystalline phase, under the condition of   
phase segregation between a DNA-rich phase and   a  
PEG-rich phase (DNA 20 $\mu $g/ml, PEG  
230 mg/ml, NaCl 100 mM).	
(c)   Polarization microscopy image  
on the  same region as in (b).
(d) Coiled DNA molecule in dilute DNA solution (DNA 0.1 $\mu $g/ml, PEG  
170 mg/ml, NaCl 100 mM).
(e) Compact DNA in dilute DNA solution (DNA 0.1 $\mu $g/ml, PEG 230  
mg/ml, NaCl 100 mM).
  (a'), (b'), (c') and (d')    
 show   fluorescence-intensity  
distributions   for (a), (b), (c) and (d),  
respectively.

FIG.3: (a)   Phase behavior of DNA-PEG solution   
represented as the volume fraction of the DNA-rich phase, the zero volume fraction corresponds to  
  a homogeneous solution
(T4 DNA 20 $\mu$g/ml, NaCl 100 mM).  Below 
of 170 mg/ml of PEG, the system exhibits homogeneous state.
Above 180 mg/ml of PEG,  
a liquid crystalline DNA phase was induced accompanied by
phase segregation.
(b)  Conformational change  of single DNA. The dependence of the  
 average long-axis length is  shown  
together with the standard deviation (T4 DNA 0.1 $\mu$g/ml, NaCl  
100 mM).
 The coiled and  
compact states  of DNA molecules coexisted at 210 mg/ml of PEG.

\clearpage

\clearpage

\hspace{5.0cm}
\clearpage

\end{document}